\documentclass[prc,aps,nofootinbib,showkeys,showpacs,preprint]{revtex4}
\usepackage{graphicx}  
\usepackage{epstopdf} 
\usepackage{dcolumn}   
\usepackage{bm}        
\usepackage{amssymb}   
\usepackage{amsmath}
\usepackage{longtable}

\begin{document}


\title{Investigation of the properties of $1^-_1$ excited states of even-even nuclei}

\author{R.V.Jolos$^{1,2}$ and E.A.Kolganova$^{1,2}$ }
\affiliation{$^1$Joint Institute for Nuclear Research, 141980 Dubna, Moscow region, Russia\\
$^2$Dubna State University, 141982 Dubna, Moscow Region, Russia}
\date{\today}

\begin{abstract}
\begin{description}
\item[Background:] The commissioning of new sources of monochromatic $\gamma$-quanta will allow obtaining a large amount of new data on low-lying collective states of negative parity of even-even nuclei.
\item[Purpose:] To derive relations between observables characterizing the low-energy negative parity states in even-even nuclei and to investigate  the dependence of E2 transitions between the low-energy collective states in these nuclei on the parity of states.
\item[Methods:] The Hamiltonian of the Collective Model and the fermionic Q-phonon representation of the vectors of collective states are used.
\item[Results:]  Relations between different E1 transition matrix elements are derived and the parity dependence of E2 transitions is investigated.
\item[Conclusion:] Relations between the observables characterizing the low-lying collective states of both parities  in even-even nuclei are obtained, which can be verified using experimental data produced by facilities generating intense monochromatic beams of $\gamma$-quanta.
\end{description}
\end{abstract}

\maketitle

\section{Introduction}

Laser Compton scattering is the only practical method for producing energy-tunable and quasi-monoenergetic photon beams at MeV energies
\cite{Litvinenko,Hajima,INOK}. Photons of MeV energies can be used to study nuclear structure because they interact with nuclei via electromagnetic interaction, which is well known.  Detecting reactions such as ($\gamma,\gamma$') is a common approach to investigate nuclear structure. Among the problems that can be investigated in these reactions, the study of the properties of the lowest lying $1^-$ states of even-even nuclei is  natural. They are excited from the ground state by dipole transition, require relatively small energies for their excitation, and are located in the low density region of nuclear excited states.

In near magic and close to them nuclei, the lowest $1^-_1$ states have a quadrupole-octupole two-phonon structure. They were observed in many nuclei of this type~\cite{Cottle,Zilges,Herzberg,Kneissl,Ibbotson,Fransen,Pietralla,Shneidman}. The conclusion about the collective nature of $1^-_1$ states was made on the basis of a number of facts \cite{Pietralla}. These include:\\
-- experimental data  on excitation energies and E1 transition probabilities which show a smooth dependence on mass number, which is typical for collective states;\\
--  excitation energies of the $1^-_1$ states which correlate with the sum  of  excitation energies of the $3^-_1$ and $2^+_1$ collective states;\\
-- E1 $0^+_1\rightarrow 1^-_1$ transitions that are characterized by relatively strong transition probabilities which are $10^3$ times stronger than typical low-energy E1 transitions.

In deformed nuclei, octupole oscillations take place in the axially deformed nuclear mean field. As a result, four negative parity states with K=0,1,2, and 3, on which rotational bands are based, appear in even-even nuclei.  Rotational bands with K=0 and 1 begin with $1^-$ states, which can be observed in experiments with  $\gamma$-beams.

A special group of nuclei is formed by the isotopes of Rn, Ra, Th, U, and Pu in which $1^-$ states with unusually small excitation energies, on which alternating parity bands are based are observed.

Below we consider a number of characteristics of the lowest $1^-$ states that are of interest for study using the Compton source of monochromatic $\gamma$-rays.

\section{Transition from nuclei with $E(1^-_1)$\textgreater $E(3^-_1)$ to nuclei with $E(1^-_1)$\textless $E(3^-_1)$}

Experimental data show that in semimagic and other nuclei with a small number of valence nucleons the excitation energies of 
the $1^-_1$ states are close to the sum of excitation energies of the $3^-_1$ and $2^+_1$ states. For example, in $^{138}$Ba
$E(1^-_1)$=4026 keV and $E(3^-_1)+E(2^+_1)=4317$ keV. In $^{142}$Nd $E(1^-_1)=3425$ keV and $E(3^-_1)+E(2^+_1)=3661$ keV.
In $^{142}$Ce $E(1^-_1)=2188$ keV and $E(3^-_1)+E(2^+_1)=2294$ keV \cite{nndc}. However, in the limit of axially deformed nuclei, the $1^-_1$ state becomes  lower in energy than the $3^-_1$ state. Thus, in transition from spherical to deformed nuclei, a dramatic change in the spectrum of excited states of negative parity occurs. If in spherical nuclei $E(1^-_1)$ \textgreater $E(3^-_1)$,
then in deformed nuclei  $E(1^-_1)$\textless $E(3^-_1)$. This fact, associated with changes in the spectrum of excited negative parity states, provides additional information about the phase transition from spherical to deformed nuclei. 
In order to study this transition, it is necessary to obtain experimental data on excited negative parity states in a variety of isotopic chains of various elements, which involve both spherical and deformed nuclei.

 To describe the change in the structure and properties of low-lying  negative parity excited states of even-even nuclei in the evolution from a spherical to a deformed shape, we use the following Hamiltonian:
\begin{eqnarray}
\label{H1}
H=H'_{quad} +\hbar\omega_3\sum_{\mu}f^+_{3\mu}f_{3\mu}-\frac{1}{2}\kappa\sum_{\mu}(-1)^{\mu}Q_{2\mu}Q_{2-\mu}.
\end{eqnarray}
Here $H'_{quad}$ depends only on quadrupole variables and takes a different form in spherical, transitional and deformed nuclei. It is assumed here that the octupole mode can be approximately treated  as a harmonic vibrator. However, the energy of the octupole phonon varies during the evolution from spherical to deformed nuclei. The quadrupole moment operator in (\ref{H1}) contains contributions of both quadrupole and octupole modes. They can be determined using the definition of $Q_{2\mu}$ in the Geometrical Collective Model:
\begin{eqnarray}
\label{Q}
Q_{2\mu}=\rho_0 \int_0^{R(\vartheta,\varphi)}d\Omega r^2Y_{2\mu}r^2dr,
\end{eqnarray} 
where
\begin{eqnarray}
\label{R}
R(\vartheta,\varphi)=R_0(1+\sum_{\mu}\alpha_{2\mu}Y^*_{2\mu}+\sum_{\mu}\alpha_{3\mu}Y^*_{3\mu}+...).
\end{eqnarray} 
As a result, we obtain
\begin{eqnarray}
\label{Qa}
Q_{2\mu}=Q_{2\mu}^{(quad)}+Q_{2\mu}^{(oct)},\nonumber\\
Q_{2\mu}^{(quad)}=\frac{3}{4\pi}AR^2_0\alpha_{2\mu}+...\nonumber\\
Q_{2\mu}^{(oct)}=\frac{3}{4\pi}AR^2_0\cdot\sqrt{\frac{28}{15\pi}}\sum_{\nu,\nu'}C^{2\mu}_{3\nu  3\nu'}\alpha_{3\nu}\alpha_{3\nu'}+...
\end{eqnarray} 
The Hamiltonian (\ref{H1}) can be written with the help of (\ref{Qa}) as \cite{Nomura,Zolnowski}:
\begin{eqnarray}
\label{H2}
H=H_{quad} +\hbar\omega_3\sum_{\mu}f^+_{3\mu}f_{3\mu}-\kappa\sum_{\mu}(-1)^{\mu}Q_{2\mu}^{(quad)}Q_{2-\mu}^{(oct)},
\end{eqnarray} 
where
\begin{eqnarray}
\label{H3}
H_{quad}=H'_{quad} -\frac{1}{2}\kappa\sum_{\mu}(-1)^{\mu}Q_{2\mu}^{(quad)}Q_{2-\mu}^{(quad)}.
\end{eqnarray}
The operator $Q_{2\mu}^{(oct)}$ can be expressed in terms of the  octupole phonon creation ($f^+_{3\mu}$) and annihilation 
($f_{3\mu}$) operators
using the following expression for $\alpha_{3\mu}$: 
\begin{eqnarray}
\label{Alpha}
\alpha_{3\mu}=\sqrt{\frac{\hbar}{2B_3\omega_3}}(f^+_{3\mu}+(-1)^{\mu}f_{3-\mu}),
\end{eqnarray} 
and neglecting by terms in $Q_{2\mu}^{(oct)}$ changing the number of octupole phonons by two, because of the
smallness of the corresponding effect:
\begin{eqnarray}
\label{Qoct}
Q_{2\mu}^{(oct)}=-\frac{3}{4\pi}AR^2_0\cdot\frac{2}{\sqrt{3\pi}}\frac{\hbar}{B_3\omega_3}\sum_{\nu,\nu'}C^{3\nu}_{3\nu'  2\mu}f^+_{3\nu}f_{3\nu'}.
\end{eqnarray}
The coefficient $\hbar$/$B_3\omega_3$ in (\ref{Qoct}) can be expressed through $B(E3;3^-_1\rightarrow 0^+_1)$ using the expression for the octupole transition operator following from the Geometrical Collective Model: 
\begin{eqnarray}
\label{Q3oct}
Q_{3\mu}=\frac{3}{4\pi}ZR_0^3\alpha_{3\mu},
\end{eqnarray} 
We obtain:
\begin{eqnarray}
\label{Eq}
\frac{\hbar}{B_3\omega_3}=2(\frac{3}{4\pi}ZR^3_0)^{-2}B(E3;3^-_1\rightarrow 0^+_1).
\end{eqnarray}

As noted above, we assume that in transition from spherical nuclei to deformed ones octupole oscillations can continue to be considered as harmonic oscillations around  a mirror symmetric, although possibly deformed, shape. However, the value of $\hbar\omega_3$ changes. Of course, this assumption excludes from consideration a number of nuclei from the region of actinides with 
low-lying $1^-_1$ states. However, in this transition, the nature of the quadrupole mode changes dramatically from being vibrational motion in the spherical limit to rotational motion.  The energy of the $2^+_1$ state goes down and it gets a quadrupole moment. Due to the coupling of the quadrupole and octupole modes, the spectrum of the low-lying negative parity states also changes.

Let us find an expression for the energy of the $1^-_1$ state $E(1^-_1)$. To calculate $E(1^-_1)$, we assume that the following relationship for the vector of the $1^-_1$ state can be used:
\begin{eqnarray}
\label{Vector5}
|1^-_1 M\rangle=\sum_{\mu,\nu}C^{1M}_{2\mu  3\nu} f^+_{3\nu}|2^+_1 \mu\rangle,
\end{eqnarray}
where the structure of the $2^+_1$ state is determined by the Hamiltonian $H_{quad}$ of (\ref{H3}). Averaging Hamiltonian (\ref{H2})
over the state (\ref{Vector5}), we obtain
\begin{eqnarray}
\label{E1Eq6}
E(1^-_1)\equiv\langle 1^-_1|H| 1^-_1\rangle=E(2^+_1)+\hbar\omega_3\nonumber\\
+\kappa\cdot \frac{8}{5}\sqrt{\frac{2}{7\pi}}\cdot\frac{3}{4\pi}AR^2_0\frac{B(E3;3^-_1\rightarrow 0^+_1)}{(\frac{3}{4\pi}ZR^3_0)^2}\nonumber\\
\times\frac{A}{Z}\langle 2^+_1 ||Q^{(\pi,quad)}_{2\mu} ||2^+_1\rangle, 
\end{eqnarray}
where $Q_{2\mu}^{(\pi,quad)}$ is the proton contribution to the mass quadrupole operator $Q_{2\mu}^{(quad)}$
Thus, due to the appearance of the quadrupole moment of the $2^+_1$ state, the energy of the $1^-_1$ state decreases if
$\langle 2^+_1 ||Q^{(quad)}_{2\mu} ||2^+_1\rangle$ is negative, as it should be in the case of prolate deformation. 

For the vector of the $3^-_1$ state we use  the  following expression as the first approximation:
\begin{eqnarray}
\label{Vector7}
| 3^-_1 M\rangle =f^+_{3M}|0^+_1\rangle.
\end{eqnarray}
 In this case, for energy of the $3^-_1$ state we get
\begin{eqnarray}
\label{E(3)8}
E(3^-_1)=\hbar\omega_3
\end{eqnarray}
Substituting this result into (\ref{E1Eq6}), we obtain the following relationship between the experimental data characterizing the
$1^-_1, 3^-_1$ and $2^+_1$ states:
\begin{eqnarray}
\label{E1Eq9}
E(1^-_1)=E(2^+_1)+E(3^-_1)\nonumber\\
+\kappa\cdot \frac{8}{5}\sqrt{\frac{2}{8\pi}}\cdot\frac{3}{4\pi}AR^2_0\frac{B(E3;3^-_1\rightarrow 0^+_1)}{(\frac{3}{4\pi}ZR^3_0)^2}\nonumber\\
\times\frac{A}{Z}\langle 2^+_1 ||Q^{(quad)}_{2\mu} ||2^+_1\rangle. 
\end{eqnarray}
Consideration of corrections to expression (\ref{Vector7}) associated with the quadrupole-octupole interaction results in a second order correction  in magnitude of $\left(E(1^-_1)-E(2^+_1)-E(3^-_1)\right)$ in (\ref{E1Eq9}), which is small.

As an example of application of relation (\ref{E1Eq9}), consider $^{148}$Nd for which the values of
$B(E3;3^-_1\rightarrow 0^+_1)=45442 e^2fm^6, \langle 2^+_1 ||Q_2 ||2^+_1\rangle =-185 e fm^2, E(2^+_1)=0.302$ MeV and 
$E(3^-_1)=0.999$ MeV are known. 
As the value of $\kappa$, we use the results of \cite{Kisslinger} where, for example, the following value was obtained for stable nuclei with Z=60-64:
\begin{eqnarray}
\label{kappa}
\kappa=\frac{0.176}{A^{4/3}} MeV\cdot fm^{-4} .
\end{eqnarray}
Substituting these values into the right-hand side of relation (\ref{E1Eq9}), we obtain
a value of 1.062 MeV for $E(1^-_1)$ while its experimental value is 1.024 MeV. Having in mind the approximations  made, we can say that the obtained result  is in qualitative agreement with the experimental data. 
 
Relation (\ref{E1Eq9}) includes the parameter $\kappa$ used in a concrete nuclear structure model. This parameter is usually a smooth function of $A$. For instance, if the nuclear mean field potential is approximated by the harmonic oscillator, then $\kappa_2\sim A^{-5/3}$ \cite{BM2}.
However, for a group of nuclei with close values of mass number $A$ this parameter can be considered as constant, for instance, for a chain of isotopes of a certain element. In this case, for nuclei from this group the expression
\begin{eqnarray}
\label{Aindependent}
\frac{E(1^-_1)-E(2^+_1)-E(3^-_1)}{1.6\sqrt{\frac{2}{8\pi}}\cdot\frac{3}{4\pi}AR^2_0\frac{B(E3;3^-_1\rightarrow 0^+_1)}{(\frac{3}{4\pi}ZR^3_0)^2}\frac{A}{Z}\langle 2^+_1 ||Q^{(quad)}_{2\mu} ||2^+_1\rangle.} 
\end{eqnarray}
is a constant. This relationship can be verified using experimental data for isotopic chains of certain elements whose amount needs to be significantly increased.

\section{Reduced E1 transition probabilities}

\subsection{Relationship between $B(E1;1^-_1\rightarrow 0^+_1)$ and $B(E1;3^-_1\rightarrow 2^+_1)$.}

To understand the properties of low-lying negative parity states, it is important to find out if there are any systematic relationships between the probabilities of E1 transitions inherent in many nuclei. Such a relationship was found between the probabilities of E1 transitions from $1^-_1$ to the ground state and from $3^-_1$ to the $2^+_1$ state \cite{Pietralla}. 

The method convenient for investigating the general relationships between the characteristics of low-lying collective states was suggested in \cite{JShV}. This is the Q-phonon approach for  fermionic configurational space. This approach is ideologically based on the Q-phonon representation of low-lying collective states obtained in the framework of the Interacting Boson Model. It was shown that the vectors of yrast, second $2^+$ and second $0^+$ states can be described with  high accuracy  over the whole parameter space of the consistent-Q Hamiltonian  by simple universal expressions containing only one or two configurations \cite{Pietralla1,Pietralla2,Palchikov}.
A simple structure of wave vectors helps derive different relations between transition matrix elements.

To derive relations between E1 transition probabilities from the $1^-_1$ state to the ground $0^+_1$ state and from $3^-_1$ to the $2^+_1$ state, we need the expressions for the vectors of the $2^+_1$, $3^-_1$, and $1^-_1$ states.  In the fermionic Q-phonon approach, the vectors of the $2^+_1$ and $3^-_1$ states can be represented as:
\begin{eqnarray}
\label{Eq18}
|2^+_1 M\rangle = {\cal N}_{2^+_1}{\hat Q}_{2M} |0^+_1\rangle,\quad {\cal N}^{-2}_{2^+_1}=\frac{1}{\sqrt{5}}\langle 0^+_1|({\hat Q}_2{\hat Q}_2)_0|0^+_1\rangle,\nonumber\\
|3^-_1 M\rangle = {\cal N}_{3^-_1}{\hat Q}_{3M} |0^+_1\rangle, \quad {\cal N}^{-2}_{3^-_1}=\frac{1}{\sqrt{7}}\langle 0^+_1|({\hat Q}_3{\hat Q}_3)_0|0^+_1\rangle,
\end{eqnarray}
where ${\cal N}_{2^+_1}$ and ${\cal N}_{3^-_1}$ are the normalization coefficients. The expression for the $1^-_1$ state vector, given by (\ref{Vector5}),  can be generalized as
\begin{eqnarray}
\label{Eq19}
|1^-_1 M\rangle ={\cal N}_{1^-_1}\sum_{\mu,\nu} C^{1 M}_{2\mu 3\nu} Q_{3\nu}|2^+_1 \mu\rangle.
\end{eqnarray}
As it is shown in \cite{JShV}, ${\cal N}_{1^-_1}\approx{\cal N}_{3^-_1}$.

Using Eqs. (\ref{Eq18}), (\ref{Eq19}) and the fact that the multipole operators $Q_{3\mu}, Q_{2\mu}$, and the E1 transition operator T(E1) commute with each other, since they depend only on the nucleons coordinates, we obtain (see Eqs. (10) and (11) in \cite{JShV}):
\begin{eqnarray}
\label{Eq20}
\langle 1^-_1 M|T(E1)_M|0^+_1\rangle ={\cal N}_{1^-_1}\sum_{\mu,\nu} C^{1 M}_{2\mu 3\nu}\langle 2^+_1\mu|{\hat Q}^+_{3\nu}T(E1)_M|0^+_1\rangle\nonumber\\
={\cal N}_{1^-_1}\sum_{\mu,\nu} C^{1 M}_{2\mu 3\nu}\langle 2^+_1\mu|T(E1)_M{\hat Q}^+_{3\nu}|0^+_1\rangle\nonumber\\
=\sum_{\mu,\nu} C^{1 M}_{2\mu 3\nu}(-1)^{\nu}\langle 2^+_1\mu|T(E1)_M|3^-_1 -\nu\rangle,
\end{eqnarray}
from which it follows that
\begin{eqnarray}
\label{Eq21}
\langle1^-_1||T(E1)||0^+_1\rangle=\langle 2^+_1||T(E1)||3^-_1\rangle.
\end{eqnarray}
Relation (\ref{Eq21}) means that
\begin{eqnarray}
\label{Eq22}
\frac{B(E1;1^-_1\rightarrow 0^+_1)}{B(E1;3^-_1\rightarrow 2^+_1)}=\frac{7}{3}.
\end{eqnarray}
This result was earlier obtained  in \cite{Pietralla}.

The derived relation (\ref{Eq22}) does not explain the observed reduction of the ratio
${B(E1;1^-_1\rightarrow 0^+_1)}/{B(E1;3^-_1\rightarrow 2^+_1)}$ to unity in a number of nuclei, for example, in Sn isotopes in which the energy of the $1^-_1$ state takes values in the range  of 3-4 MeV. It can not be excluded that the $1^-_1$ state contains an impurity of a component of a different nature than the quadrupole-octupole component (\ref{Eq19}). This circumstance may be the reason for the deviation of the experimental value of the ratio ${B(E1;1^-_1\rightarrow 0^+_1)}/{B(E1;3^-_1\rightarrow 2^+_1)}$ from the theoretical prediction.

\subsection{The branching ratio $R\equiv B(E1;1^-_1\rightarrow 2^+_1)$/$B(E1;1^-_1\rightarrow 0^+_1)$}

In deformed nuclei, the branching ratio $R=\frac{B(E1;1^-_1\rightarrow 2^+_1)}{B(E1;1^-_1\rightarrow 0^+_1)}$ makes it possible to determine the value of the projection of the total angular momentum of the $1^-_1$ state on the symmetry axis. According to the Alaga rules,
$R=2$ in the case of $K=0$ and $R=0.5$ in the case of $K=1$. Below we show  that in the case of transitional nuclei, the branching ratio $R$ correlates with the quadrupole moment of the $1^-_1$ state.

Let us consider the matrix element $\langle 1^-_1 M'|T(E1)_{\mu}| 2^+_1 M\rangle$ using the Q-phonon representation  of the $2^+_1$ state vector from (\ref{Eq18}) and the commutativity of the operators T(E1) and $Q_2$. We obtain
\begin{eqnarray}
\label{Eq31a}
\langle 1^-_1 M'|T(E1)_{\mu}| 2^+_1 M\rangle ={\cal N}_{2^+_1}\langle 1^-_1M'|Q_{2M}T(E1)_{\mu}|0^+_1\rangle.
\end{eqnarray}
It follows from (\ref{Eq31a}) that
\begin{eqnarray}
\label{Eq32a}
\langle 1^-_1 M'|T(E1)_{\mu}| 2^+_1 M\rangle ={\cal N}_{2^+_1}\sum_n\langle 1^-_1M'|Q_{2M}|1^-_n\mu\rangle\langle 1^-_n\mu|T(E1)_{\mu}|0^+_1\rangle.
\end{eqnarray}

In the case of spherical and transitional nuclei, there is only one strong E1 transition from the ground state to relatively low-lying states. Using this fact, we can get from(\ref{Eq32a}) the approximate relation
\begin{eqnarray}
\label{Eq32}
\langle 1^-_1 M'|T(E1)_{\mu}| 2^+_1 M\rangle \approx{\cal N}_{2^+_1}\langle 1^-_1M'|Q_{2M}|1^-_1\mu\rangle\langle 1^-_1\mu|T(E1)_{\mu}|0^+_1\rangle.
\end{eqnarray}
Note, however, that the summation over intermediate $1^-$ states in (\ref{Eq32a}) includes, in principle, the contribution of states that form the Giant Dipole Resonance. And even if the wave function of the $1^-_1$ state includes a small contribution of this state, the small amount of the corresponding contribution to $\langle 1^-_1M'|Q_{2M}|1^-_{GDR}\mu\rangle$ can be compensated by the large value of the matrix element $\langle 1^-_{GDR}\mu|T(E1)_{\mu}|0^+_1\rangle$. This circumstance can introduce an error into the relation obtained below for the quadrupole moment of the $1^-_1$ state. 

From (\ref{Eq32}) it follows that 
\begin{eqnarray}
\label{Eq33}
\langle 1^-_1||T(E1)||2^+_1\rangle =\frac{1}{\sqrt{3}}{\cal N}_{2^+_1}\langle 1^-_1||Q_2|| 1^-_1\rangle\langle 1^-_1||T(E1)||0^+_1\rangle
\end{eqnarray}
Using (\ref{Eq33}), (\ref{Eq18}) and the definition of reduced transition probabilities in terms of the reduced matrix elements, we obtain
\begin{eqnarray}
\label{Eq33a}
|\langle 1^-_1||Q_2||1^-_1\rangle| =\sqrt{3B(E2;2^+_1\rightarrow 0^+_1)\frac{B(E1;1^-_1\rightarrow 2^+_1)}{B(E1;1^-_1\rightarrow 0^+_1)}}
\end{eqnarray}
As an example, consider again $^{148}$Nd for which $B(E2;2^+_1\rightarrow 0^+_1), B(E1;1^-_1\rightarrow 2^+_1)$ and
$B(E1;1^-_1\rightarrow 0^+_1)$ are known \cite{Ibbotson}. Substituting these values into (\ref{Eq33a}), we obtain that
$|\langle 1^-_1||Q_2||1^-_1\rangle|$=118 $e\cdot fm^2$. The experimental value of $\langle 1^-_1||Q_2||1^-_1\rangle$
is \cite{Ibbotson}: -39$^{+40}_{-8}$ $e\cdot fm^2$.

The expression for $\langle 1^-_1||Q_2||1^-_1\rangle$ can be obtained directly using expression (\ref{Eq19}) for the wave vector of the $1^-_1$ state and (\ref{Qa}), (\ref{Qoct}) and (\ref{Eq}) for the $Q_{2\mu}$ operator. As a result, we obtain
\begin{eqnarray}
\label{Eq35}
\langle 1^-_1||Q_2||1^-_1\rangle = -\frac{16\sqrt{6\pi}}{15}\frac{A B(E3;3^-_1\rightarrow 0^+_1)}{Z^2R^4_0}+\frac{1}{5}\sqrt{\frac{3}{7}}\langle 2^+_1||Q_2||2^+_1\rangle,
\end{eqnarray}
where $R_0$ is the nuclear radius, and A and Z are the nuclear mass and charge numbers. Substituting  the experimental data for $^{148}$Nd \cite{Ibbotson} into (\ref{Eq35}), we obtain that $\langle 1^-_1||Q_2||1^-_1\rangle=-29 e\cdot fm^2$, which is in agreement with the experimental value. The contribution of the first term in (\ref{Eq35}) is equal to $-5e\cdot fm^2$.

\subsection{Dependence of E2 transition probability on parity of states between which transitions occur}

Experimental studies of low-lying states of negative parity in actinides and heavy Ba and Ce isotopes have shown that these nuclei have a mirror asymmetric shape or, at the very least, are characterized by strong mirror asymmetric correlations and a highly collectivized quadrupole mode \cite{Ahmad,Butler}. Experimental data on E2 transition probabilities between states forming alternating parity bands in $144$Ba show that E2 transitions between negative parity states are significantly stronger than those between positive parity ones \cite{Shneidman}. This effect can be explained by a larger value of the quadrupole deformation in negative parity states compared to positive parity states. This situation can be realized if the potential energy of a nucleus in the plane of $\beta_{20} - \beta_{30}$ variables has two minima symmetrically located at $\beta_{30}\neq 0$. The barrier that separates these two minima is situated at $\beta_{30}=0$ and is distinguished by a smaller quadrupole deformation than both minima.  Due to the fact that the wave functions of negative parity states are odd with respect to the change of the sign of $\beta_{30}$, these wave functions are equal to zero at the barrier and shifted towards the minima of potential energy, i.e. towards greater quadrupole deformation. The wave functions of positive parity states are even with respect to the change of the sign of $\beta_{30}$ and are non-zero at the barrier. As the result, they have effectively smaller quadrupole deformation than the negative parity states. The results of calculation of the potential energy of $^{218,220,226}$Ra \cite{Unzhakova} support a possibility of such interpretation.

A completely  different picture is realized in nuclei, which, in contrast to $^{144}$Ba, do not have  octupole deformation and  are not even characterized by large amplitude octupole oscillations.  In these nuclei, octupole oscillations occur relative to 
$\beta_{30}=0$. 
In this case,  the physical picture can be analyzed based on the Hamiltonian (\ref{H2}). If  the  nuclei under consideration have axially symmetric quadrupole deformation, then the Hamiltonian (\ref{H2}) can be transformed into an intrinsic frame by:
\begin{eqnarray}
Q^{(quad)}_{2\mu}\rightarrow D^2_{\mu 0}Q_0,\nonumber\\
f^+_{3\mu}\rightarrow \sum_K D^3_{\mu K}f^+_{3 K},
\end{eqnarray}
and the collective momentum of the quadrupole mode $R_{\mu}$ is transformed as
\begin{eqnarray}
R_{\mu}\rightarrow I_{\mu}-j^{(oct)}_{\mu},
\end{eqnarray}
where $I_{\mu}$ is the total angular momentum and $j^{(oct)}_{\mu}$ is the angular momentum of the octupole mode. Assuming that
$H_{quad}=\frac{\hbar^2}{2\Im}{\vec R}^2$, we obtain \cite{Shneidman}:
\begin{eqnarray}
\label{H19}
H=\frac{\hbar^2}{2\Im}({\vec I}-{\vec j}^{(oct)})^2+\sum_K(\hbar\omega_3+\kappa Q_0 C^{3K}_{3K 20})f^+_{3K}f_{3K}.
\end{eqnarray}
Due to the Coriolis interaction, the eigenfunctions of single-phonon states with negative parity, unlike states with positive parity, are K-mixed and represented by
\begin{eqnarray}
\label{EF20}
\sqrt{\frac{2I+1}{8\pi^2}} \sum_Ku^{(I)}_KD^I_{M K}f^+_{3K}|0^+_1\rangle,
\end{eqnarray}
where the coefficients $u^{(I)}_K$ are determined by the diagonalization of the Hamiltonian (\ref{H19}). In this case,  the parity of the states $\pi$ is related to the angular momentum as $\pi=(-1)^I$.
It was shown in 
\cite{Shneidman} that in the considered case:
\begin{eqnarray}
\label{Q21}
\langle I||Q_2||I-2\rangle=\sqrt{\frac{5}{16\pi^2}}Q_0\sqrt{2I-3}C^{I 0}_{I-2 0   20}\nonumber\\
\times\left(1-\frac{1}{2}(1-(-1)^I)\frac{\langle I(I-2)|K^2|I(I-2)\rangle}{(I-\frac{1}{2})^2}\right),
\end{eqnarray}
where $\langle I(I-2)|K^2|I(I-2)\rangle=\sum_K\left(u^{(I(I-2))}_K\right)^2K^2$. It is seen from (\ref{Q21}) that in the case of even I (positive parity), the expression in the parenthesis in (\ref{Q21}) is equal to unity. However, in the case of odd I (negative parity), this expression is smaller than unity. this means that the E2 reduced transition matrix elements for positive parity states are larger than for negative parity states.  It is indeed the case in $^{148}$Nd \cite{Ibbotson}.

\section{Conclusion}

It is shown that using fermionic Q-phonon representations of low-lying collective excitations of both parities in even-even nuclei, several relations between observables can be derived. 
To verify the correctness of these ratios, a sufficient amount of new experimental data is required.
New facilities that produce energy-tunable, quasi-monoenergetic photon beams with MeV energies will increase the amount of data. The parity dependence of the strength of E2 transitions between states of the lowest rotational bands is discussed.

\acknowledgments{
The authors are grateful to the National Center for Physics and Mathematics (Russia) for support.
}



\begin{thebibliography}{99}

\bibitem{Litvinenko} P.G.O'Shea, V.N.Litvinenko, J.M.J.Madey \textit{et al.}, Nuclear Instruments and Methods in Physics Research Section A, {\bf 375}, 530 (1996).
\bibitem{Hajima} R.Hajima, Physics Procedia {\bf 84}, 35 (2016).
\bibitem{INOK} L.V.Grigorenko et al., FyzMat 1, 121 (2023).
\bibitem{Cottle} P.D.Cottle and D.A.Bromley, Phys.Lett. B \textbf{182}, 129 (1986).
\bibitem{Zilges} A.Zilges, P. von Brentano, H.Friedrichs, R.D.Heil, U.Kneissl, S. Lindenstruth, H.H.Pitz, and C.Wesselborg, Z.Phys. A \textbf{340}, 155 (1991).
\bibitem{Herzberg} R.-D.Herzberg, I.Bauske, P. von Brentano, Th. Eckert, R.Fisher, W. Geiger, U. Kneissl, J. Margraf, H.Maser, N. Pietralla, H.H. Pitz, A. Zilges, Nucl.Phys. A \textbf{592}, 211 (1995).
\bibitem{Kneissl} U.Kneissl, H.H.Pitz, and A. Zilges, Prog.Part.Nucl.Phys.\textbf{37}, 349 (1996).
\bibitem{Ibbotson} R.W. Ibbotson, C.A.White, T. Czosnyka, P.A. Butler, N. Clarkson, D.Cline et al. Nucl.Phys. A \textbf{619}, 213 (1997).
\bibitem{Fransen} C.Fransen, O.Beck, P. von Brentano, T.Eckert, R.-D. Herzberg, U. Kneissl, H. Maser, A. Nord, N. Pietralla, H.H. Pitz, and A. Zilges, Phys.Rev. C \textbf{57}, 129 (1998). 
\bibitem{Pietralla} N. Pietralla, Phys.Rev. C  \textbf{59}, 2941 (1999).
\bibitem{Shneidman} T.M.Shneidman, R.V. Jolos, R. Kr\"ucken, A. Aprahamian, D. Cline, J.R.Cooper, M. Cromaz, R.M. Clark, C. Hutter, A.O. Machiavelli, W. Scheid, M.A. Stoyer, and C.Y. Wu, Eur.Phys. J. A \textbf{25}, 387 (2005).
\bibitem{nndc} NNDC, Evaluated Nuclear Structure Data File, http://www.nndc.bnl.gov/ensdf/.
\bibitem{Nomura} M.Nomura, Phys.Lett. B {\bf 55}, 357 (1975).
\bibitem{Zolnowski} D.R.Zolnowski, T.Kishimoto, Y.Gono, and T.T.Sugihara, Phys.Lett. B {\bf 55}, 453 (1975). 
\bibitem{Kisslinger} L.S.Kisslinger, R.A.Sorensen, Rev.Mod.Phys. {\bf 35}, 853 (1963).
\bibitem{BM2} A.Bohr and B.R.Mottelson, {\it Nuclear Structure} (Benjamin, Reading, 1985), Vol. 2.
\bibitem{JShV} R.V.Jolos, N.Yu.Shirikova, V.V.Voronov, Phys.Rev. C {\bf 70}, 054303 (2004).
\bibitem{Pietralla1} N.Pietralla, P. von Brentano, R.F.Casten, T.Otsuka, andN.V.Zamfir, Phys.Rev.Lett. {\bf 73}, 2962 (1994).
\bibitem{Pietralla2} N.Pietralla, P. von Brentano, T.Otsuka, and R.F.Casten, Phys.Lett. B {\bf 349}, 1 (1995).
\bibitem{Palchikov} Yu.V.Palchikov, P. von Brentano, R.V.Jolos, Phys.Rev. C {\bf 57}, 3026 (1998).
\bibitem{Ahmad} I.Ahmad, P.A.Butler, Ann.Rev.Nucl.Part.Sci. {\bf 43}, 71 (1993).
\bibitem{Butler} P.A.Butler, W.Nazarewicz, Rev.Mod.Phys. {\bf 68}, 349 (1996).
\bibitem{Unzhakova} R.V.Jolos, Yu.V.Palchikov, V.V.Pashkevich, and A.V.Unzhakova, Nuovo Cim. {\bf 110A}, 941 (1997).


\end{thebibliography}
\end{document}